\documentclass[aps,prb,twocolumn,superscriptaddress,showpacs]{revtex4}
\usepackage{graphicx}
\usepackage{latexsym}
\usepackage{amssymb}
\usepackage{amsmath}
\usepackage{amsfonts}
\usepackage{bm}
\usepackage{multirow}
\usepackage{color}
\usepackage{comment}
\newcommand{\ii}{\mathrm{i}}

\renewcommand{\Im}{\mathop{\mathrm{Im}}}

\newcommand{\U}{\mathrm{U}}

\newcommand{\beq}{\begin{equation}}
\newcommand{\eeq}{\end{equation}}
\newcommand{\beqn}{\begin{eqnarray}}
\newcommand{\eeqn}{\end{eqnarray}}

\DeclareMathAlphabet{\mathbbold}{U}{bbold}{m}{n}

\def\sgn{{\rm sgn}}

\def\U{{\rm U}}

\begin{document}

\title{Orbital Orders and Possible non-Fermi Liquid in Moir\'{e} systems}

\author{Yichen Xu}
\affiliation{Department of Physics, University of California,
Santa Barbara, CA 93106, USA}

\author{Xiao-Chuan Wu}
\affiliation{Department of Physics, University of California,
Santa Barbara, CA 93106, USA}

\author{Chao-Ming Jian}
\affiliation{Station Q, Microsoft, Santa Barbara, California
93106-6105, USA}

\author{Cenke Xu}
\affiliation{Department of Physics, University of California,
Santa Barbara, CA 93106, USA}

\begin{abstract}

Motivated by recent observation of nematicity in Moir\'{e}
systems, we study three different orbital orders that potentially
can happen in Moir\'{e} systems: (1) the nematic order; (2) the
valley polarization; and (3) the ``compass order". Each order
parameter spontaneously breaks part of the spatial symmetries of
the system. We explore physics caused by the quantum fluctuations
close to the order-disorder transition of these order parameters.
Especially, we recognize that the symmetry of the Moir\'{e}
systems leads to a crucial difference of the effective theory
describing the nematic order from the standard Hertz-Millis
formalism. We demonstrate that this key difference may lead to a
special non-Fermi liquid behavior near the order-disorder nematic
transition, different from the standard non-Fermi liquid behavior
usually expected when a Fermi surface is coupled to the critical
fluctuations of orbital orders. We also discuss the interplay of
the three order parameters and the possible rich phase diagram at
finite temperature. Within the three orbital orders, the valley
polarization and the ``compass order" likely strongly compete with
the superconductor.

\end{abstract}

\maketitle

\section{Introduction}

Systems with Moir\'{e} superlattice have surprised the condensed
matter community with a plethora of correlated phenomena,
supposedly due to the strong Coulomb interaction and the
narrowness of the minibands in the Moir\'{e} mini Brillouin
zone~\cite{flat1,flat2,flat3,flat4,TDBG1,TDBG2,TDBG3}. Correlated
insulator at fractional fillings~\cite{wangmoire,mag01}, high
temperature superconductor (compared with the miniband
width)~\cite{mag02,young2018,TLGSC,FM,kimtalk,efetov,TDBG1,TDBG2,TDBG3},
quantum anomalous Hall effect~\cite{zaletel,zhangmao,hall,hall2},
strange metal (non-Fermi liquid)~\cite{moirestrange,youngstrange},
competing orders~\cite{pablotalk,pablonematic}, spin-triplet
pairing~\cite{TDBG1,TDBG2,TDBG3,TDBGt} have all been reported in
recent experiments on Moir\'{e} systems. Many of these phenomena
may have to do with order parameters with nontrivial
transformations under spatial symmetries, $i.e.$ the orbital
orders. For example, the quantum anomalous Hall effect definitely
requires valley polarization because the Chern numbers of two
degenerate minibands from two different valleys must cancel each
other due to symmetry~\cite{zaletel,zhangmao,hall,hall2}. Also,
strong signature of nematic anisotropy was found in recent
experiments on twisted bilayer graphene, in both the
superconductor phase and the metallic
phase~\cite{pablotalk,pablonematic}. Mean field analysis of
orbital orders in lattice models related to Moir\'{e} systems have
also been studied~\cite{kivelsontbg}.

Motivated by the experimental observations, in this work we
discuss possible orbital orders in Moir\'{e} systems. We will
explore novel generic physics at the order-disorder transition of
the orbital orders, based on the spatial symmetries of the
systems. Three different kinds of orbital orders, $i.e.$ (1) the
nematic order, (2) valley polarization, and (3) ``compass order",
which spontaneously break different subgroups of the entire
spatial symmetries will be discussed. These orders should be
viewed as possible instability of Fermi surface due to
interactions. We will focus on the order-disorder quantum phase
transition of these order parameters, and especially how the
quantum fluctuations of these order parameters may affect the
electrons. We demonstrate that, due to the unique symmetry of the
systems, the nematic order fluctuation may lead to a special
non-Fermi liquid behavior, different from what is usually expected
at the quantum critical regime of an orbital order. The interplay
between these order parameters allows a very rich phase diagram at
zero and finite temperature. Within these three orbital orders,
the valley polarization and ``compass order" can potentially
strongly compete with the superconductor.

\section{Three orbital orders}

In all the Moir\'{e} systems discovered so far, the most general
microscopic symmetry is $C_3 \times \mathcal{T}$, where
$\mathcal{T}$ is an effective time-reversal symmetry which is a
product between the ordinary time-reversal and a spin-flipping,
hence this effective time-reversal symmetry still holds even with
a background Zeeman field (inplane magnetic field). Under this
symmetry, the Fermi surface of the miniband emerging from each
valley only has a $C_3$ symmetry, and $\mathcal{T}$ interchanges
the two valleys. The dispersion of the minibands from the two
valleys satisfy $\varepsilon_1(\vec{k}) = \varepsilon_2(-
\vec{k})$, where the subscript is the valley index. Different
Moir\'{e} systems have different extra symmetries, for example the
twisted bilayer graphene (TBG) without alignment with the BN
substrate has an inversion symmetry $\mathcal{I}$, while the
trilyer graphene and h-BN heterostructure has a reflection
symmetry $\mathcal{P}$~\cite{senthil}. Both $\mathcal{I}$ and
$\mathcal{P}$ interchange the two
valleys~\cite{senthil,band1,band4}. We assume that the system
under study has the symmetry $C_3 \times \mathcal{T} \times
\mathcal{I}$. Under these spatial symmetries, the momenta and
electron operators transform as \beqn C_3 &:& (k_x + \ii k_y)
\rightarrow e^{\ii 2\pi/3} (k_x + \ii k_y); \cr\cr \mathcal{T}&:&
c_{a,\vec{k}} \rightarrow \tau^1_{ab} c_{b,-\vec{k}}, \ \
\mathcal{I}: c_{a,\vec{k}} \rightarrow \tau^1_{ab} c_{b,
-\vec{k}}, \eeqn where $a,b$ are the valley indices. In this work
we will discuss three different orbital orders, each breaking
different subgroups of the entire symmetry $C_3 \times \mathcal{T}
\times \mathcal{I}$.

The first orbital order we will consider is the nematic order
$\phi$, which is a complex scalar order parameter. The microscopic
operator of the nematic order parameter in a two dimensional
($2d$) rotational invariant system can be written
as~\cite{fradkinnematic} \beqn \hat{\phi}(\vec{x}) \sim
\psi^\dagger (\vec{x}) ( \partial_x^2 -
\partial_y^2 + \ii 2 \partial_x \partial_y) \psi(\vec{x}), \eeqn
where $\psi(x)$ is the real space electron operator. $\hat{\phi}$
is an operator with zero or small momentum compared with the Fermi
wave vector. In a system with symmetry $C_3 \times \mathcal{T}
\times \mathcal{I}$, the zero momentum nematic operator can be
represented as \beqn \hat{\phi} &\sim& \sum_{\vec{k}}
c^\dagger_{1,\vec{k}}\left( k_x^2 - k_y^2 + 2 \ii k_x k_y + \alpha
( k_x - \ii k_y ) \right) c_{1,\vec{k}} \cr\cr &+& \sum_{\vec{k}}
c^\dagger_{2,\vec{k}}\left( k_x^2 - k_y^2 + 2 \ii k_x k_y - \alpha
( k_x - \ii k_y ) \right) c_{2,\vec{k}} \label{nematic} \eeqn with
real number $\alpha$. Since the Fermi surface on each valley only
has a $C_3$ symmetry, the $d_{x^2 - y^2} +\ii d_{xy}$ order
parameter with angular momentum $(+2)$ will mix with a $p_x - \ii
p_y$ order parameter with angular momentum $(-1)$. The nematic
order parameter $\phi \sim \langle \hat{\phi} \rangle$ transforms
under the symmetries as \beqn C_3 : \phi \rightarrow e^{\ii
2\pi/3} \phi; \ \ \mathcal{T} : \phi \rightarrow \phi^\ast, \ \
\mathcal{I} : \phi \rightarrow \phi. \eeqn A nonzero condensate of
$\phi$ will break the spatial symmetries down to $\mathcal{T}$ and
$\mathcal{I}$ only, and in this sense we can still refer to $\phi$
as a nematic order parameter. Nematic order has been found in many
condensed matter systems (for a review see
Ref.~\onlinecite{nematicreview}), and strong signature of the
existence of nematic order in both the superconducting phase and
the normal metallic phase was recently reported in
TBG~\cite{pablotalk,pablonematic}.

The second orbital order we will discuss is the valley
polarization $\Phi$, which corresponds to an operator \beqn
\hat{\Phi} \sim \sum_{\vec{k}} c^\dagger_{1,\vec{k}}c_{1,\vec{k}}
- c^\dagger_{2,\vec{k}} c_{2,\vec{k}} . \eeqn A valley
polarization $\Phi \sim \langle \hat{\Phi} \rangle$ is an Ising
like order parameter. A nonzero $\Phi$ will cause imbalance of the
electron density between the two valleys, $i.e.$ the electron has
higher population at one valley than the other, and it may lead to
the quantum anomalous Hall
effect~\cite{zaletel,zhangmao,hall,hall2}. $\Phi$ preserves the
$C_3$ symmetry, but breaks both $\mathcal{T}$ and $\mathcal{I}$.

The last order parameter is the ``compass order" which is again a
complex scalar order parameter. The microscopic compass order
operator is represented as \beqn \hat{\varphi} &\sim&
\sum_{\vec{k}} c^\dagger_{1,\vec{k}}\left( k_x^2 - k_y^2 + 2 \ii
k_x k_y + \alpha ( k_x - \ii k_y ) \right) c_{1,\vec{k}} \cr\cr
&-& \sum_{\vec{k}} c^\dagger_{2,\vec{k}}\left( k_x^2 - k_y^2 + 2
\ii k_x k_y - \alpha ( k_x - \ii k_y ) \right) c_{2,\vec{k}}.
\eeqn Under the symmetry actions, the compass order parameter
$\varphi \sim \langle \hat{\varphi} \rangle$ transforms as \beqn
C_3 : \varphi \rightarrow e^{\ii 2\pi/3} \varphi; \ \ \mathcal{T}:
\varphi \rightarrow - \varphi^\ast, \ \ \mathcal{I}: \varphi
\rightarrow - \varphi. \eeqn The symmetry transformation of
$\varphi$ can be viewed as the definition of the order parameter.
$\varphi$ also has the same symmetry transformation as the
composite field $\phi \Phi$.

The full symmetry $C_3 \times \mathcal{T} \times \mathcal{I}$
guarantees that, a nonzero nematic order leads to three different
degenerate ground states, while a compass order can take six
different expectation values with degenerate energy. The compass
order and valley polarization both break time-reversal symmetry
$\mathcal{T}$, hence both orders can lead to anomalous Hall
effect, as was observed in Ref.~\onlinecite{hall,hall2}. Since the
nematic order preserves $\mathcal{T}$, a nematic order alone
cannot lead to the anomalous Hall signal. But a nematic order
breaks the rotation symmetry, hence it directly couples to the
background strain of the system.

\section{Order-Disorder Transition of the Nematic Order}

Normally when an order parameter with zero or small momentum
couples to the Fermi surface, the dynamics of the order parameter
is over-damped at low frequency according to the standard
Hertz-Millis theory~\cite{hertz,millis}. The nematic order
parameter is slightly more complicated, when coupled to a circular
Fermi surface, the dynamics of the nematic order parameter is
decomposed into a transverse mode and longitudinal mode, and only
the longitudinal mode is over-damped. The separation of the two
modes was computed explicitly in Ref.~\onlinecite{fradkinnematic},
whose physical picture can be understood as following. Consider a
general order parameter with a small momentum $\vec{q}$, the
over-damping of this mode comes from its coupling with the patch
of Fermi surface where the tangential direction is parallel with
$\vec{q}$. For a circular Fermi surface, without loss of
generality, let us assume $\vec{q} = (q_x, 0)$, then the Fermi
patches that cause over-damping locate at $\vec{k}_f \sim \pm
\hat{y}$. But $\Im[\phi]$ defined previously has nodes along the
$\pm \hat{y}$ direction (rotational invariance guarantees that the
``tangential patch" of the Fermi surface coincides with the node
of the transverse mode), hence $\Im[\phi]_{\vec{q}}$ with $\vec{q}
= (q_x, 0)$ is not over-damped.

But now the symmetry of the system, especially the fact that the
$d-$wave order parameter mixes with the $p-$wave order parameter,
no longer guarantees that for any small momentum $\vec{q}$ the
``tangential patch" of the Fermi surface coincides with the node
of the order parameter, hence $\phi$ is always over-damped, which
can be shown with explicit calculations following
Ref.~\onlinecite{fradkinnematic}. Thus we will start with the
following Hertz-Millis type of action for the nematic order
parameter $\phi$, which is invariant under the symmetry $C_3
\times \mathcal{T} \times \mathcal{I}$: \beqn \mathcal{S}_b &=&
\mathcal{S}_0 + \int d^2x d\tau \ u (\phi^3 + \phi^{\ast 3}) + g
|\phi|^4, \cr\cr \mathcal{S}_0 &=& \sum_{\vec{q}, \omega}
\phi^\ast_{\vec{q},\omega} \left( \frac{|\omega|}{q} + q^2 + r
\right) \phi_{\vec{q},\omega} \label{action} \eeqn For convenience
we have written the free part of the action $\mathcal{S}_0$ in the
momentum and Matsubara frequency space, but the interaction terms
of the action in the Euclidean space-time. Also, since the $\U(1)$
rotation of $\phi$ is in fact a spatial rotation, there should be
coupling between the direction of $\vec{\phi} =
(\mathrm{Re}[\phi], \mathrm{Im}[\phi])$ and direction of momentum,
which we have ignored for simplicity~\footnote{The simplest
nonzero term of this type would be $\int d^2x d\tau \phi
(\partial_x - \ii
\partial_y)^2 \phi$, which has the same scaling as the rest of the
quadratic terms, hence it is not expected to lead to more singular
contribution to the physical quantities to be calculated.}.
Following the standard Hertz-Millis theory~\cite{hertz,millis},
the action Eq.~\ref{action} is scaling invariant if we assign the
following scaling dimensions to the parameters and field: \beqn
[\omega] = 3, \ \ [q_x] = [q_y] = 1, \ \ [r] = 2, \cr\cr
[\phi(\vec{x},\tau)] = \frac{3}{2}, \ \ [u] = \frac{1}{2}, \ \ [g]
= -1. \label{scaling} \eeqn At the level of the Hertz-Millis
theory, normally the total space-time dimension is greater than
the upper critical dimension, hence the self-interaction of the
order parameter is usually irrelevant, and the theory will lead to
an ordinary mean field transition (for a review see
Ref.~\onlinecite{review}). However, unlike the ordinary
Hertz-Millis theory, in our current case there is an extra
symmetry-allowed term $u (\phi^3 + \phi^{\ast 3})$ that is
relevant even though the total space-time dimension is $D = d + z
= 5$. Thus we need to perform analysis beyond the mean field
theory, and explore the possible new physics led by the new term.

The relevant $u$ term breaks the $\U(1)$ symmetry of $\phi$ down
to a $Z_3$ symmetry, which is the symmetry of a three-state clock
model~\cite{clock}. A mean field analysis of such Ginzburg-Landau
theory would lead to a first order transition which occurs at $r_c
= u^2/g$, but a two dimensional three-state clock model
(equivalent to a three-state Potts model) has a continuous
transition and can be potentially described by the Ginzburg-Landau
theory with a $Z_3$ anisotropy on a $\U(1)$ order
parameter~\cite{6epsilon}. Ref.~\onlinecite{yao2017,bisenthil2}
also presented examples of first order quantum phase transitions
at the mean field level (precisely due to a cubic term like our
$u-$term in the action) being driven to continuous transitions by
fluctuations, especially when the order parameter is coupled to
gapless fermions~\cite{yao2017}, which is analogous to our
situation.

Without knowing for sure the true nature of the transition
described by Eq.~\ref{action}, at least the scaling analysis in
the previous paragraph applies when $r$ is tuned {\it close to}
while {\it greater than} $r_c$, and in the energy scale $\omega
\gg (u^2/g)^{3/2}$ the order parameter $\phi$ can always be viewed
as a massless scalar field with self-interaction $u$ and $g$ in
Eq.~\ref{action}.

If we further assume that $(u^2/g)^{3/2} \ll 1/g^3$ and only look
at energy scale $\omega < 1/g^3$, the irrelevant coupling $g$ is
renormalized small enough. Hence when the parameters in
Eq.~\ref{action} satisfy $(u^2/g)^{3/2} \ll 1/g^3$, there is a
finite energy window $\omega \in \left( (u^2/g)^{3/2}, \ 1/g^3
\right)$ where we can view $\phi$ as a massless scalar field which
interacts with itself mainly through the $u$ term in
Eq.~\ref{action}, and the ordinary $|\phi|^4$ interaction is
irrelevant and renormalized perturbatively weak. We expect that
the action Eq.~\ref{action} with the relevant interaction $u$ can
lead to new universal physics that is beyond the standard
Hertz-Millis theory.

\begin{figure}
\includegraphics[width=0.65\linewidth]{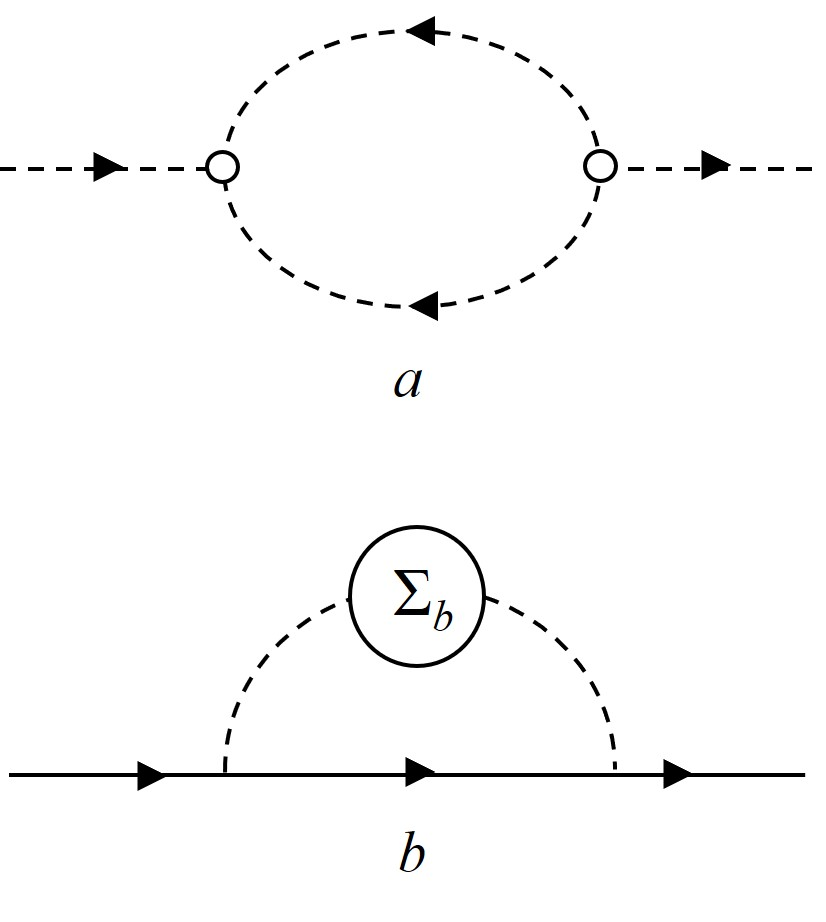}
\caption{ $a$, the one-loop correction to the boson propagator
from the $u$ term in Eq.~\ref{action}; $b$, the one-loop
correction (Eq.~\ref{marginal}) to the fermion propagator through
the boson-fermion coupling $g'$ in Eq.~\ref{action2}.} \label{fd1}
\end{figure}

Based on Eq.~\ref{action}, if we take into account of the relevant
perturbation $u$, in general the boson propagator reads \beqn
G_b(\omega, \vec{q}) &=& \frac{1}{G^{-1}_{b0}(\omega, \vec{q}) +
\Sigma_b(\omega, \vec{q})}, \cr\cr G^{-1}_{b0}(\omega, \vec{q})
&=& \frac{|\omega|}{q} + q^2. \eeqn A full reliable analysis of
Eq.~\ref{action} with the relevant perturbation $u$ is difficult,
we will first limit our study to the lowest nontrivial order of
perturbation of $u$, later we will discuss other analysis. At the
one-loop level (Fig.~\ref{fd1}$a$), the boson self-energy
$\Sigma_b(\omega, \vec{p})$ reads \beqn \Sigma_b(\omega,\vec{q})
&\sim& u^2 \int d^2k d\nu \ G_{b0}(\nu, \vec{k}) G_{b0}(\omega +
\nu, \vec{q} + \vec{k}), \cr\cr &\sim& \mathrm{Const} + A u^2
\sqrt{|\omega|^{2/3} + c q^2} + \cdots \label{Gb}\eeqn $A$ and $c$
are both order one constants. The behavior of the boson
self-energy is consistent with power-counting of the loop
integral, and at low energy it dominates other quadratic terms
$\mathcal{S}_0$ in the standard Hertz-Millis theory, due to the
fact that $u$ is a relevant perturbation. The cut-off dependent
constant can be reabsorbed into $r$, and the ellipsis includes
terms that are less dominant in the infrared~\footnote{The loop
integral is performed numerically, and it fits best with the
expression in Eq.~\ref{Gb}.}.

For our purpose we need to analyze the effects of the
boson-fermion coupling on the electrons. In the standard
Hertz-Millis theory without the relevant $u$ term in the boson
action, the one loop self-energy of the electron scales as
$\Sigma_f(\omega) \sim \ii \mathrm{sgn}[\omega] |\omega|^{2/3}$.
We will analyze how the $u$ term may change the behavior of the
fermion self-energy. Following the formalism used in
Ref.~\onlinecite{nayak1,nayak2,mrossnfl,sslee,maxnematic}, we
expand the system at one patch of the Fermi surface. The
``one-patch" theory is a very helpful formalism to systematically
evaluate loop diagrams in a boson-fermion coupled theory. This
``one-patch theory" breaks the $C_3$ symmetry, hence the real and
imaginary parts of $\phi$ are no longer degenerate. Since we are
most interested in the scaling behavior of the Fermion
self-energy, we will consider a one component boson field with the
dressed propagator and self-energy given by Eq.~\ref{Gb}. The
one-patch theory reads \beqn \mathcal{S}_{bf} &=& \sum_{\omega,
\vec{k}} \ \psi^\dagger_{\omega, \vec{k}}(\ii \omega - v_f k_x - v
k_y^2)\psi_{\omega, \vec{k}} \cr\cr &+& \mathcal{S}_0 +
\sum_{\omega, \vec{q}} \Sigma_b (\omega, \vec{q})
|\phi_{\omega,\vec{q}}|^2 \cr\cr &+& \int d^2x d\tau \ g' \phi
\psi^\dagger \psi, \label{action2}\eeqn Where $\mathcal{S}_0$ is
given by Eq.~\ref{action}, and $\Sigma_b(\omega,\vec{q})$ given by
Eq.~\ref{Gb}.

For this ``one-patch" boson-fermion coupled theory we need to use
a different assignment of scaling dimensions, which was introduced
in Ref.~\onlinecite{nayak1,nayak2,mrossnfl,sslee,maxnematic} for a
better controlled analysis of the boson-fermion coupled theory. In
order to avoid confusion, we use $``[ \ ]"$ to denote the scaling
dimension of the original pure boson theory Eq.~\ref{action}, but
$``\{ \ \}"$ to denote the scaling dimension of the ``one-patch"
boson-fermion coupled theory: \beqn \{\omega\} = 3, \ \ \{k_x\} =
2, \ \ \{k_y\} = 1, \cr\cr \{\phi(\vec{x},\tau)\} = \frac{5}{2}, \
\ \{\psi\} = 2, \ \ \{g'\} = - \frac{1}{2}. \label{scaling2}\eeqn
Under the new scaling relation Eq.~\ref{scaling2}, $\mathcal{S}_0$
becomes irrelevant compared with $\Sigma_b(\omega, \vec{q})$ in
Eq.~\ref{Gb}. We will first ignore the irrelevant term
$\mathcal{S}_0$ completely (which will be revisited later) to
reveal the main effect of the new $u-$term in Eq.~\ref{action}.
The one-loop fermion self-energy (Fig.~\ref{fd1}$b$) reads \beqn
\Sigma_f(\omega) &\sim& \int d^2k d\nu \ G_{f0}(\nu, \vec{k})
G_b(\omega + \nu, \vec{k}) \cr\cr &\sim& \int d^2k d\nu
\frac{1}{\ii \nu - v_f k_x - v k_y^2} \frac{1}{\sqrt{|\omega +
\nu|^{2/3} + c k^2}} \cr\cr &\sim& \ii \omega \log \left(
\frac{\Lambda}{|\omega|} \right). \label{marginal}\eeqn This
behavior of fermion self-energy is similar to the marginal fermi
liquid, and it is consistent with the simple power-counting of the
loop integral. The marginal fermi liquid was proposed as a
phenomenological theory for the strange metal phase (a non-Fermi
liquid phase) of the cuprates high temperature
superconductor~\cite{mfl}. A similar strange metal behavior was
observed in the TBG~\cite{pablotalk,moirestrange,youngstrange}.
Our goal here is not to directly address the observed strange
metal bahavior~\footnote{It was suggested that a pure
phonon-electron coupling can lead to a linear$-T$ resistivity in
Moir\'{e} systems~\cite{phononstrange,youngstrange}, but it was
argued in Ref.~\onlinecite{moirestrange} that other mechanisms may
be demanded to explain the observed data.}, instead we stress that
the electrons at the order-disorder transition of the nematic
order in the Moir\'{e} systems should behave differently from what
is usually expected at a nematic quantum critical point. This
difference originates from the unique symmetry of the Moir\'{e}
systems.

Because $g'$ is an irrelevant perturbation in Eq.~\ref{action2}
according to the scaling convention of the ``one-patch" theory
Eq.~\ref{scaling2}, higher order perturbation of $g'$ in theory
Eq.~\ref{action2} is not expected to lead to more dominant
correction to the fermion self-energy in the infrared, hence we no
long need to worry about the infinite ``planar diagram" problem in
ordinary cases when an order parameter is coupled with a Fermi
surface~\cite{sslee}.

\begin{figure}
\includegraphics[width=0.65\linewidth]{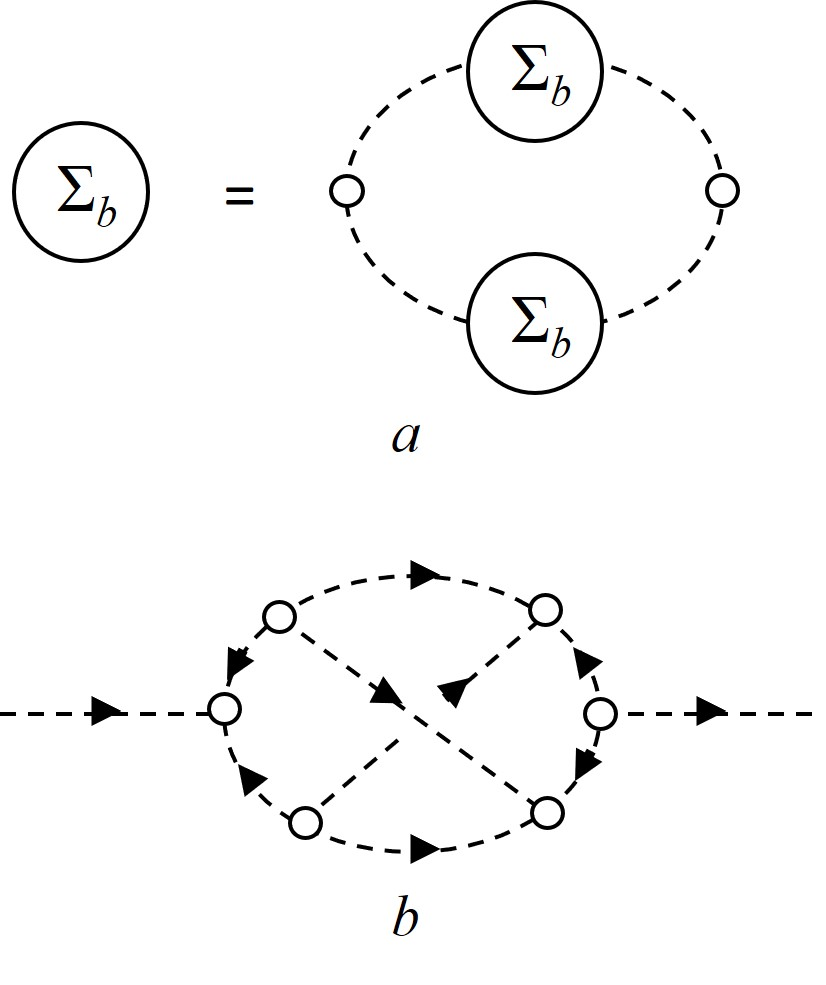}
\caption{ $a$, the schematic representation of the Schwinger-Dyson
equation; $b$, the example of vertex correction that is not summed
in the Schwinger-Dyson equation. } \label{fd2}
\end{figure}

The results above are based on the one-loop calculation in the
expansion of $u$, and higher order expansion of $u$ will modify
the results in the infrared limit. If eventually the nematic
transition is driven continuous by fluctuations as the examples
given in Ref.~\onlinecite{yao2017,bisenthil2}, then a full
analysis for the infrared limit is desired. Although we cannot
completely solve the strongly interacting theory Eq.~\ref{action}
analytically beyond the perturbation expansion, an approximate
solution can be obtained through the Schwinger-Dyson (SD)
equation, which sums a subset of the Feynman diagrams
Fig.~\ref{fd2}$a$: \beqn \Sigma_b &\sim& u^2 \int d^2k d\nu \
G_{b}(\nu, \vec{k}) G_{b}(\omega + \nu, \vec{q} + \vec{k}), \cr\cr
G_b^{-1} &=& G_{b0}^{-1} + \Sigma_b. \eeqn Here we have ignored
the vertex correction from the full SD equation (For example,
vertex correction Fig.~\ref{fd2}$b$). We also take a simple ansatz
that at the order-disorder transition the boson self-energy is
approximated by the scaling form \beqn \Sigma_b(\omega, \vec{q})
\sim u^{2\eta} Q^{2 - \eta} \eeqn with anomalous dimension $\eta$,
where $Q$ is the infrared cut-off that can be taken as
$\mathrm{Max}[|\omega|^{1/3}, |\vec{q}|]$. The previous one-loop
result simply yields $\eta = 1$. Now the bosonic part of the
boson-fermion coupling action Eq.~\ref{action2} is replaced by $
\mathcal{S}_b = \mathcal{S}_0 + \sum_{\omega, \vec{p}}
\Sigma_b(\omega, \vec{q}) |\phi_{\omega, \vec{q}}|^2$. Again, if
we tentatively ignore $\mathcal{S}_0$, the scaling of the
boson-fermion coupling theory is modified as \beqn \{\omega\} = 3,
\ \ \{k_x\} = 2, \ \ \{k_y\} = 1, \cr\cr \{\phi(\vec{x},\tau)\} =
\frac{4 + \eta}{2}, \ \ \{\psi\} = 2, \ \ \{g'\} = -
\frac{\eta}{2}. \eeqn The one-loop fermion self-energy should then
scale as \beqn \Sigma_f(\omega) \sim \ii \omega
|\omega|^{\frac{\eta - 1}{3}}. \label{fself}\eeqn As long as $\eta
> 0$, the boson-fermion coupling $g'$ in Eq.~\ref{action2} is still
irrelevant, hence higher order fermion self-energy diagrams from
the boson-fermion coupling theory are not expected to change
Eq.~\ref{fself} in the infrared.

The solution of the approximate SD equation would yield $\eta =
1/3$, which will lead to a non-fermi liquid behavior that is
in-between the standard Hertz-Millis theory and also the marginal
fermi liquid. After we convert the Matsubara frequency to real
frequency, the imaginary part of the fermion self-energy (inverse
of the fermion life-time) is a very characteristic property of the
non-Fermi liquid. And the analysis above suggests that the
imaginary part of fermion self-energy should scale as
$\mathrm{Im}(\Sigma_f) \sim \sgn(\omega)|\omega|^\beta$ with $2/3
< \beta < 1$.

If eventually the transition in Eq.~\ref{action} is driven
continuous by fluctuation (like the examples given in
Ref.~\onlinecite{yao2017,bisenthil2}), then the field $\phi$ is
indeed massless even in the infrared limit at the transition. Then
the difference of our results described above from the standard
Hertz-Millis theory should be obvious in the infrared limit. But
even if the transition is first order at $r_c = u^2/g$, as we
discussed previously, when the parameters in Eq.~\ref{action}
satisfy $(u^2/g)^{3/2} \ll 1/g^3$, at least there is a finite
energy window $|\omega| \in \left( (u^2/g)^{3/2}, \ 1/g^3 \right)$
where $\phi$ can be viewed as a massless scalar field with strong
self-interaction mainly through the $u$ term, while other
interactions in Eq.~\ref{action} can be ignored. In this case, the
calculations in this section were simplified by assuming that the
boson self-energy $\Sigma_b$ dominates the other quadratic term,
$i.e.$ $\mathcal{S}_0$ in Eq.~\ref{action}, because
$\mathcal{S}_0$ is irrelevant compared with $\Sigma_b$. But in the
finite energy window described above, since we are not in the
infrared limit, one should keep a nonzero $\mathcal{S}_0$ together
with $\Sigma_b$ in the calculation of the fermion self-energy.
Then the loop integral in the evaluation of $\Sigma_f(\omega)$ is
more complicated. The fermion self-energy is no longer a simple
scaling form $\mathrm{Im}(\Sigma_f) \sim
\sgn(\omega)|\omega|^\beta$ with a constant exponent $\beta$.
Instead, the exponent $\beta$ is expected to increase from the
standard Hertz-Millis result $\beta = 2/3$ while decreasing
$\omega$, $i.e.$ in other words the system should crossover back
to the standard Hertz-Millis result at higher energy scale. We
have numerically calculated the fermion self-energy by keeping a
nonzero $\mathcal{S}_0$ in the bosonic theory, and confirmed this
expectation of crossover.

Recently the standard result of the fermion self-energy scaling of
the Hertz-Millis theory was confirmed in numerical
simulation~\cite{meng23} on nematic transitions on a square
lattice. We expect that our qualitative prediction of the fermion
self-energy under the symmetry of the Moir\'{e} systems can also
be seen in future numerical simulations.

\section{Valley Polarization and Compass order}

The effective theory of the valley polarization order $\Phi$ and
compass order $\varphi$ are more conventional Hertz-Millis
theories whose analysis can be quoted from
Ref.~\onlinecite{review}. A cubic self-interaction term is not
allowed for either order parameters. But the symmetry
transformation of the compass order $\varphi$ allows a term \beqn
u_6 (\varphi^6 + \varphi^{\ast 6})\eeqn in the
Ginzburg-Landau-Hertz-Millis theory of $\varphi$, which is
irrelevant in the infrared at the total space-time dimension $D =
5$. The three order parameters are coupled together in the
effective theory, and the lowest order symmetry-allowed couplings
are: \beqn \mathcal{L}_{mix} &=& \cdots + r_\phi |\phi^2| +
r_{\varphi} |\varphi|^2 + r_\Phi |\Phi|^2 \cr\cr &+& v_{1} (\Phi
\phi \varphi^{\ast} + h.c.) + v_2 (\Phi \varphi^3 + h.c.) \cr\cr
&+& v_3 \Phi^2 |\phi|^2 + v_4 \Phi^2 |\varphi|^2. \label{action3}
\eeqn A full exploration of the multi-dimensional parameter space
will lead to a very complex and rich phase diagram. The specific
values of the parameters in Eq.~\ref{action3} depend heavily on
the microscopic physics of the system.

\begin{figure}
\includegraphics[width=\linewidth]{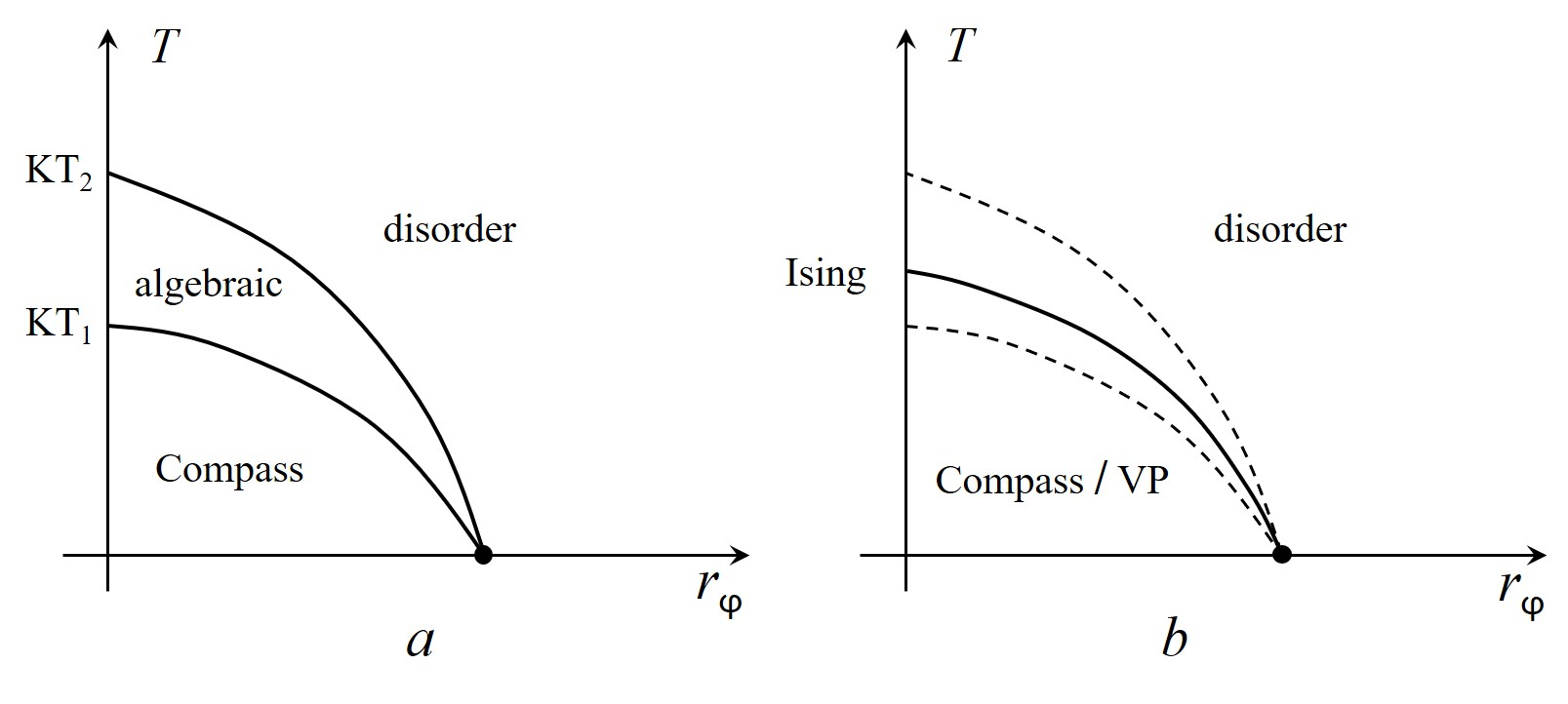}
\caption{ $a$, the phase diagram when there is a compass order at
zero temperature, there are two consecutive Kosterlitz-Thouless
transitions at finite temperatures and an algebraic phase in
between; $b$, once there is a background strain in the system, the
compass order is identical to the valley polarization (VP) order,
and hence there is only one Ising transition at finite
temperature.} \label{pd}
\end{figure}

Recently evidence of strain that breaks the $C_3$ rotation
symmetry has been reported in Moir\'{e} systems~\cite{strain}, and
the strain can potentially strongly affect the band
structure~\cite{strainband}. With a background strain field, the
nematic order parameter $\phi$ acquires a nonzero expectation
value, and hence $\Phi$ and $\varphi$ become the same order
parameter through the coupling $v_1$ in $\mathcal{L}_{mix}$.

At finite temperature, the nematic order and valley polarization
will go through continuous transitions which correspond to the
three-state potts and Ising conformal field theory with central
charges $4/5$ and $1/2$ respectively. While if we start with a
zero temperature compass order, the finite temperature physics can
be mapped to a six-state clock model due to the $u_6$ term
mentioned previously in the Ginzburg-Landau theory of the compass
order. In this case while raising temperature the system will
undergo two consecutive continuous Kosterlitz-Thouless transitions
with an algebraic quasi-long range order in between. Within the
algebraic phase, the scaling dimension of the compass order
parameter $[\varphi]$ is temperature dependent, and $ 1/18 <
[\varphi] < 1/8$. The nematic order parameter $\phi \sim
\varphi^{\ast 2}$ and valley polarization $\Phi \sim \varphi^3 +
\varphi^{\ast 3}$ also have power-law correlation function in the
algebraic phase, and their scaling dimensions are $ [\phi] = 4
[\varphi] $, and $[\Phi] = 9 [\varphi]$. Hence even a weak
background strain which pins $\phi$ is always a relevant
perturbation in the algebraic phase, and will collapse the two
Kosterlitz-Thouless transitions of $\varphi$ into a single Ising
transition of $\Phi$.

Signature of a hidden order which strongly competes with
superconductor was observed
experimentally~\cite{pablotalk,pablonematic}. Within the three
orbital orders that we have discussed in this work, the valley
polarization and compass order both obviously compete with the
superconductor. The reason is that both these two order parameters
break $\mathcal{T}$ and $\mathcal{I}$, hence break the degeneracy
between electrons at $\vec{k}$ and $-\vec{k}$ (the $C_3$ symmetry
alone does not protect this degeneracy), hence a nonzero $\Phi$ or
$\varphi$ makes it difficult to form zero momentum Cooper pair.
Indeed, experiments so far have not found superconductivity near
the quantum anomalous Hall state in Moir\'{e} systems which at
least require either valley polarization or the compass order. We
stress that the competing order mentioned here does not
necessarily mean it is the nature of the correlated insulator
observed experimentally.

\section{Final remarks}

In this work we studied three different orbital orders that may
occur in Moir\'{e} systems. We demonstrate that at the
order-disorder transition of the nematic order parameter (one of
the three orbital orders), a special non-Fermi liquid behavior is
expected in a finite energy window, due to the symmetry of the
system. We focused on the metallic phase at the disorder-order
transition of the orbital order, since experimentally a nematic
metallic phase was observed~\cite{pablonematic} above the nematic
superconducting phase. Since the three different orbital orders
can interact with each other in the effective theory and lead to a
complex and rich phase diagram, depending on the parameters the
Moir\'{e} systems under different conditions may display different
orbital orders. We demonstrate that the effective theory for the
nematic order is beyond the standard Hertz-Millis theory.
Numerical methods such as
Ref.~\onlinecite{nematicnumerics1,nematicnumerics2,nematicnumerics3,meng23}
are demanded to verify the results in the current work.

We focused on the generic field theory analysis of the phase
transitions of the orbital orders, based on the symmetry of the
system. The parameters of the field theory can be estimated
through a calculation based on the lattice models, but this
estimate depends on the microscopic details of the systems, and it
may vary strongly between different systems. For example, the
parameter $u$ stems from the $C_3$ symmetry of the Fermi surface
at each valley, and its value depends on the extent of the $C_3$
deformation from the ordinary circular Fermi surface, which likely
strongly depends on the microscopic model as well as the charge
density. Due to the complexity and subtlety of the microscopic
analysis, we plan to leave it to future studies. In the future we
will also pursue a proper generalized renormalization group
expansion such as Ref.~\onlinecite{nayak1,nayak2,mrossnfl}, as
well as analysis of the stability of the nematic order transition
towards other orders such as
superconductivity~\cite{nflpair1,nflpair2,nflpair3,nflinstable} in
Moir\'{e} systems.

The authors thank Leon Balents, Steve Kivelson for helpful
discussions. This work is supported by NSF Grant No. DMR-1920434,
the David and Lucile Packard Foundation, and the Simons
Foundation.

\bibliography{nematic}

\begin{thebibliography}{54}
\expandafter\ifx\csname natexlab\endcsname\relax\def\natexlab#1{#1}\fi
\expandafter\ifx\csname bibnamefont\endcsname\relax
  \def\bibnamefont#1{#1}\fi
\expandafter\ifx\csname bibfnamefont\endcsname\relax
  \def\bibfnamefont#1{#1}\fi
\expandafter\ifx\csname citenamefont\endcsname\relax
  \def\citenamefont#1{#1}\fi
\expandafter\ifx\csname url\endcsname\relax
  \def\url#1{\texttt{#1}}\fi
\expandafter\ifx\csname urlprefix\endcsname\relax\def\urlprefix{URL }\fi
\providecommand{\bibinfo}[2]{#2}
\providecommand{\eprint}[2][]{\url{#2}}

\bibitem[{\citenamefont{Bistritzer and MacDonald}(2011)}]{flat1}
\bibinfo{author}{\bibfnamefont{R.}~\bibnamefont{Bistritzer}} \bibnamefont{and}
  \bibinfo{author}{\bibfnamefont{A.~H.} \bibnamefont{MacDonald}},
  \bibinfo{journal}{Proceedings of the National Academy of Sciences}
  \textbf{\bibinfo{volume}{108}}, \bibinfo{pages}{12233}
  (\bibinfo{year}{2011}), ISSN \bibinfo{issn}{0027-8424},
  \eprint{http://www.pnas.org/content/108/30/12233.full.pdf},
  \urlprefix\url{http://www.pnas.org/content/108/30/12233}.

\bibitem[{\citenamefont{Su\'arez~Morell
  et~al.}(2010)\citenamefont{Su\'arez~Morell, Correa, Vargas, Pacheco, and
  Barticevic}}]{flat2}
\bibinfo{author}{\bibfnamefont{E.}~\bibnamefont{Su\'arez~Morell}},
  \bibinfo{author}{\bibfnamefont{J.~D.} \bibnamefont{Correa}},
  \bibinfo{author}{\bibfnamefont{P.}~\bibnamefont{Vargas}},
  \bibinfo{author}{\bibfnamefont{M.}~\bibnamefont{Pacheco}}, \bibnamefont{and}
  \bibinfo{author}{\bibfnamefont{Z.}~\bibnamefont{Barticevic}},
  \bibinfo{journal}{Phys. Rev. B} \textbf{\bibinfo{volume}{82}},
  \bibinfo{pages}{121407} (\bibinfo{year}{2010}),
  \urlprefix\url{https://link.aps.org/doi/10.1103/PhysRevB.82.121407}.

\bibitem[{\citenamefont{Fang and Kaxiras}(2016)}]{flat3}
\bibinfo{author}{\bibfnamefont{S.}~\bibnamefont{Fang}} \bibnamefont{and}
  \bibinfo{author}{\bibfnamefont{E.}~\bibnamefont{Kaxiras}},
  \bibinfo{journal}{Phys. Rev. B} \textbf{\bibinfo{volume}{93}},
  \bibinfo{pages}{235153} (\bibinfo{year}{2016}),
  \urlprefix\url{https://link.aps.org/doi/10.1103/PhysRevB.93.235153}.

\bibitem[{\citenamefont{Trambly~de Laissardi\`ere
  et~al.}(2012)\citenamefont{Trambly~de Laissardi\`ere, Mayou, and
  Magaud}}]{flat4}
\bibinfo{author}{\bibfnamefont{G.}~\bibnamefont{Trambly~de Laissardi\`ere}},
  \bibinfo{author}{\bibfnamefont{D.}~\bibnamefont{Mayou}}, \bibnamefont{and}
  \bibinfo{author}{\bibfnamefont{L.}~\bibnamefont{Magaud}},
  \bibinfo{journal}{Phys. Rev. B} \textbf{\bibinfo{volume}{86}},
  \bibinfo{pages}{125413} (\bibinfo{year}{2012}),
  \urlprefix\url{https://link.aps.org/doi/10.1103/PhysRevB.86.125413}.

\bibitem[{\citenamefont{Shen et~al.}(2020)\citenamefont{Shen, Chu, Wu, Li,
  Wang, Zhao, Tang, Liu, Tian, Watanabe et~al.}}]{TDBG1}
\bibinfo{author}{\bibfnamefont{C.}~\bibnamefont{Shen}},
  \bibinfo{author}{\bibfnamefont{Y.}~\bibnamefont{Chu}},
  \bibinfo{author}{\bibfnamefont{Q.}~\bibnamefont{Wu}},
  \bibinfo{author}{\bibfnamefont{N.}~\bibnamefont{Li}},
  \bibinfo{author}{\bibfnamefont{S.}~\bibnamefont{Wang}},
  \bibinfo{author}{\bibfnamefont{Y.}~\bibnamefont{Zhao}},
  \bibinfo{author}{\bibfnamefont{J.}~\bibnamefont{Tang}},
  \bibinfo{author}{\bibfnamefont{J.}~\bibnamefont{Liu}},
  \bibinfo{author}{\bibfnamefont{J.}~\bibnamefont{Tian}},
  \bibinfo{author}{\bibfnamefont{K.}~\bibnamefont{Watanabe}},
  \bibnamefont{et~al.}, \bibinfo{journal}{Nature Physics}
  (\bibinfo{year}{2020}), ISSN \bibinfo{issn}{1745-2481},
  \urlprefix\url{http://dx.doi.org/10.1038/s41567-020-0825-9}.

\bibitem[{\citenamefont{Liu et~al.}(2019)\citenamefont{Liu, Hao, Khalaf, Lee,
  Watanabe, Taniguchi, Vishwanath, and Kim}}]{TDBG2}
\bibinfo{author}{\bibfnamefont{X.}~\bibnamefont{Liu}},
  \bibinfo{author}{\bibfnamefont{Z.}~\bibnamefont{Hao}},
  \bibinfo{author}{\bibfnamefont{E.}~\bibnamefont{Khalaf}},
  \bibinfo{author}{\bibfnamefont{J.~Y.} \bibnamefont{Lee}},
  \bibinfo{author}{\bibfnamefont{K.}~\bibnamefont{Watanabe}},
  \bibinfo{author}{\bibfnamefont{T.}~\bibnamefont{Taniguchi}},
  \bibinfo{author}{\bibfnamefont{A.}~\bibnamefont{Vishwanath}},
  \bibnamefont{and} \bibinfo{author}{\bibfnamefont{P.}~\bibnamefont{Kim}}
  (\bibinfo{year}{2019}), \eprint{1903.08130}.

\bibitem[{\citenamefont{Cao et~al.}(2019)\citenamefont{Cao, Rodan-Legrain,
  Rubies-Bigorda, Park, Watanabe, Taniguchi, and Jarillo-Herrero}}]{TDBG3}
\bibinfo{author}{\bibfnamefont{Y.}~\bibnamefont{Cao}},
  \bibinfo{author}{\bibfnamefont{D.}~\bibnamefont{Rodan-Legrain}},
  \bibinfo{author}{\bibfnamefont{O.}~\bibnamefont{Rubies-Bigorda}},
  \bibinfo{author}{\bibfnamefont{J.~M.} \bibnamefont{Park}},
  \bibinfo{author}{\bibfnamefont{K.}~\bibnamefont{Watanabe}},
  \bibinfo{author}{\bibfnamefont{T.}~\bibnamefont{Taniguchi}},
  \bibnamefont{and}
  \bibinfo{author}{\bibfnamefont{P.}~\bibnamefont{Jarillo-Herrero}}
  (\bibinfo{year}{2019}), \eprint{1903.08596}.

\bibitem[{\citenamefont{Chen et~al.}(2019{\natexlab{a}})\citenamefont{Chen,
  Jiang, Wu, Lyu, Li, Chittari, Watanabe, Taniguchi, Shi, Jung
  et~al.}}]{wangmoire}
\bibinfo{author}{\bibfnamefont{G.}~\bibnamefont{Chen}},
  \bibinfo{author}{\bibfnamefont{L.}~\bibnamefont{Jiang}},
  \bibinfo{author}{\bibfnamefont{S.}~\bibnamefont{Wu}},
  \bibinfo{author}{\bibfnamefont{B.}~\bibnamefont{Lyu}},
  \bibinfo{author}{\bibfnamefont{H.}~\bibnamefont{Li}},
  \bibinfo{author}{\bibfnamefont{B.~L.} \bibnamefont{Chittari}},
  \bibinfo{author}{\bibfnamefont{K.}~\bibnamefont{Watanabe}},
  \bibinfo{author}{\bibfnamefont{T.}~\bibnamefont{Taniguchi}},
  \bibinfo{author}{\bibfnamefont{Z.}~\bibnamefont{Shi}},
  \bibinfo{author}{\bibfnamefont{J.}~\bibnamefont{Jung}}, \bibnamefont{et~al.},
  \bibinfo{journal}{Nature Physics} \textbf{\bibinfo{volume}{15}},
  \bibinfo{pages}{237每241} (\bibinfo{year}{2019}{\natexlab{a}}), ISSN
  \bibinfo{issn}{1745-2481},
  \urlprefix\url{http://dx.doi.org/10.1038/s41567-018-0387-2}.

\bibitem[{\citenamefont{Cao et~al.}(2018{\natexlab{a}})\citenamefont{Cao,
  Fatemi, Demir, Fang, Tomarken, Luo, Sanchez-Yamagishi, Watanabe, Taniguchi,
  Kaxiras et~al.}}]{mag01}
\bibinfo{author}{\bibfnamefont{Y.}~\bibnamefont{Cao}},
  \bibinfo{author}{\bibfnamefont{V.}~\bibnamefont{Fatemi}},
  \bibinfo{author}{\bibfnamefont{A.}~\bibnamefont{Demir}},
  \bibinfo{author}{\bibfnamefont{S.}~\bibnamefont{Fang}},
  \bibinfo{author}{\bibfnamefont{S.~L.} \bibnamefont{Tomarken}},
  \bibinfo{author}{\bibfnamefont{J.~Y.} \bibnamefont{Luo}},
  \bibinfo{author}{\bibfnamefont{J.~D.} \bibnamefont{Sanchez-Yamagishi}},
  \bibinfo{author}{\bibfnamefont{K.}~\bibnamefont{Watanabe}},
  \bibinfo{author}{\bibfnamefont{T.}~\bibnamefont{Taniguchi}},
  \bibinfo{author}{\bibfnamefont{E.}~\bibnamefont{Kaxiras}},
  \bibnamefont{et~al.}, \bibinfo{journal}{Nature}
  \textbf{\bibinfo{volume}{556}}, \bibinfo{pages}{80}
  (\bibinfo{year}{2018}{\natexlab{a}}).

\bibitem[{\citenamefont{Cao et~al.}(2018{\natexlab{b}})\citenamefont{Cao,
  Fatemi, Fang, Watanabe, Taniguchi, Kaxiras, and Jarillo-Herrero}}]{mag02}
\bibinfo{author}{\bibfnamefont{Y.}~\bibnamefont{Cao}},
  \bibinfo{author}{\bibfnamefont{V.}~\bibnamefont{Fatemi}},
  \bibinfo{author}{\bibfnamefont{S.}~\bibnamefont{Fang}},
  \bibinfo{author}{\bibfnamefont{K.}~\bibnamefont{Watanabe}},
  \bibinfo{author}{\bibfnamefont{T.}~\bibnamefont{Taniguchi}},
  \bibinfo{author}{\bibfnamefont{E.}~\bibnamefont{Kaxiras}}, \bibnamefont{and}
  \bibinfo{author}{\bibfnamefont{P.}~\bibnamefont{Jarillo-Herrero}},
  \bibinfo{journal}{Nature} \textbf{\bibinfo{volume}{556}}, \bibinfo{pages}{43}
  (\bibinfo{year}{2018}{\natexlab{b}}).

\bibitem[{\citenamefont{Yankowitz et~al.}(2019)\citenamefont{Yankowitz, Chen,
  Polshyn, Zhang, Watanabe, Taniguchi, Graf, Young, and Dean}}]{young2018}
\bibinfo{author}{\bibfnamefont{M.}~\bibnamefont{Yankowitz}},
  \bibinfo{author}{\bibfnamefont{S.}~\bibnamefont{Chen}},
  \bibinfo{author}{\bibfnamefont{H.}~\bibnamefont{Polshyn}},
  \bibinfo{author}{\bibfnamefont{Y.}~\bibnamefont{Zhang}},
  \bibinfo{author}{\bibfnamefont{K.}~\bibnamefont{Watanabe}},
  \bibinfo{author}{\bibfnamefont{T.}~\bibnamefont{Taniguchi}},
  \bibinfo{author}{\bibfnamefont{D.}~\bibnamefont{Graf}},
  \bibinfo{author}{\bibfnamefont{A.~F.} \bibnamefont{Young}}, \bibnamefont{and}
  \bibinfo{author}{\bibfnamefont{C.~R.} \bibnamefont{Dean}},
  \bibinfo{journal}{Science} \textbf{\bibinfo{volume}{363}},
  \bibinfo{pages}{1059} (\bibinfo{year}{2019}), ISSN \bibinfo{issn}{0036-8075},
  \eprint{http://science.sciencemag.org/content/363/6431/1059.full.pdf},
  \urlprefix\url{http://science.sciencemag.org/content/363/6431/1059}.

\bibitem[{\citenamefont{Chen et~al.}(2019{\natexlab{b}})\citenamefont{Chen,
  Sharpe, Gallagher, Rosen, Fox, Jiang, Lyu, Li, Watanabe, Taniguchi
  et~al.}}]{TLGSC}
\bibinfo{author}{\bibfnamefont{G.}~\bibnamefont{Chen}},
  \bibinfo{author}{\bibfnamefont{A.~L.} \bibnamefont{Sharpe}},
  \bibinfo{author}{\bibfnamefont{P.}~\bibnamefont{Gallagher}},
  \bibinfo{author}{\bibfnamefont{I.~T.} \bibnamefont{Rosen}},
  \bibinfo{author}{\bibfnamefont{E.~J.} \bibnamefont{Fox}},
  \bibinfo{author}{\bibfnamefont{L.}~\bibnamefont{Jiang}},
  \bibinfo{author}{\bibfnamefont{B.}~\bibnamefont{Lyu}},
  \bibinfo{author}{\bibfnamefont{H.}~\bibnamefont{Li}},
  \bibinfo{author}{\bibfnamefont{K.}~\bibnamefont{Watanabe}},
  \bibinfo{author}{\bibfnamefont{T.}~\bibnamefont{Taniguchi}},
  \bibnamefont{et~al.}, \bibinfo{journal}{Nature}
  \textbf{\bibinfo{volume}{572}}, \bibinfo{pages}{215每219}
  (\bibinfo{year}{2019}{\natexlab{b}}), ISSN \bibinfo{issn}{1476-4687},
  \urlprefix\url{http://dx.doi.org/10.1038/s41586-019-1393-y}.

\bibitem[{\citenamefont{Sharpe et~al.}(2019)\citenamefont{Sharpe, Fox, Barnard,
  Finney, Watanabe, Taniguchi, Kastner, and Goldhaber-Gordon}}]{FM}
\bibinfo{author}{\bibfnamefont{A.~L.} \bibnamefont{Sharpe}},
  \bibinfo{author}{\bibfnamefont{E.~J.} \bibnamefont{Fox}},
  \bibinfo{author}{\bibfnamefont{A.~W.} \bibnamefont{Barnard}},
  \bibinfo{author}{\bibfnamefont{J.}~\bibnamefont{Finney}},
  \bibinfo{author}{\bibfnamefont{K.}~\bibnamefont{Watanabe}},
  \bibinfo{author}{\bibfnamefont{T.}~\bibnamefont{Taniguchi}},
  \bibinfo{author}{\bibfnamefont{M.~A.} \bibnamefont{Kastner}},
  \bibnamefont{and}
  \bibinfo{author}{\bibfnamefont{D.}~\bibnamefont{Goldhaber-Gordon}},
  \bibinfo{journal}{Science} \textbf{\bibinfo{volume}{365}},
  \bibinfo{pages}{605每608} (\bibinfo{year}{2019}), ISSN
  \bibinfo{issn}{1095-9203},
  \urlprefix\url{http://dx.doi.org/10.1126/science.aaw3780}.

\bibitem[{\citenamefont{Kim}(2019)}]{kimtalk}
\bibinfo{author}{\bibfnamefont{P.}~\bibnamefont{Kim}},
  \emph{\bibinfo{title}{{Ferromagnetic superconductivity in twisted double
  bilayer graphene}}},
  \bibinfo{howpublished}{\url{http://online.kitp.ucsb.edu/online/bands_m19/kim%
/}} (\bibinfo{year}{2019}), \bibinfo{note}{{T}alks at KITP, Jan 15, 2019}.

\bibitem[{\citenamefont{Lu et~al.}(2019)\citenamefont{Lu, Stepanov, Yang, Xie,
  Aamir, Das, Urgell, Watanabe, Taniguchi, Zhang et~al.}}]{efetov}
\bibinfo{author}{\bibfnamefont{X.}~\bibnamefont{Lu}},
  \bibinfo{author}{\bibfnamefont{P.}~\bibnamefont{Stepanov}},
  \bibinfo{author}{\bibfnamefont{W.}~\bibnamefont{Yang}},
  \bibinfo{author}{\bibfnamefont{M.}~\bibnamefont{Xie}},
  \bibinfo{author}{\bibfnamefont{M.~A.} \bibnamefont{Aamir}},
  \bibinfo{author}{\bibfnamefont{I.}~\bibnamefont{Das}},
  \bibinfo{author}{\bibfnamefont{C.}~\bibnamefont{Urgell}},
  \bibinfo{author}{\bibfnamefont{K.}~\bibnamefont{Watanabe}},
  \bibinfo{author}{\bibfnamefont{T.}~\bibnamefont{Taniguchi}},
  \bibinfo{author}{\bibfnamefont{G.}~\bibnamefont{Zhang}},
  \bibnamefont{et~al.}, \bibinfo{journal}{Nature}
  \textbf{\bibinfo{volume}{574}}, \bibinfo{pages}{653每657}
  (\bibinfo{year}{2019}), ISSN \bibinfo{issn}{1476-4687},
  \urlprefix\url{http://dx.doi.org/10.1038/s41586-019-1695-0}.

\bibitem[{\citenamefont{Bultinck et~al.}(2020)\citenamefont{Bultinck,
  Chatterjee, and Zaletel}}]{zaletel}
\bibinfo{author}{\bibfnamefont{N.}~\bibnamefont{Bultinck}},
  \bibinfo{author}{\bibfnamefont{S.}~\bibnamefont{Chatterjee}},
  \bibnamefont{and} \bibinfo{author}{\bibfnamefont{M.~P.}
  \bibnamefont{Zaletel}}, \bibinfo{journal}{Physical Review Letters}
  \textbf{\bibinfo{volume}{124}} (\bibinfo{year}{2020}), ISSN
  \bibinfo{issn}{1079-7114},
  \urlprefix\url{http://dx.doi.org/10.1103/PhysRevLett.124.166601}.

\bibitem[{\citenamefont{Zhang et~al.}(2019)\citenamefont{Zhang, Mao, and
  Senthil}}]{zhangmao}
\bibinfo{author}{\bibfnamefont{Y.-H.} \bibnamefont{Zhang}},
  \bibinfo{author}{\bibfnamefont{D.}~\bibnamefont{Mao}}, \bibnamefont{and}
  \bibinfo{author}{\bibfnamefont{T.}~\bibnamefont{Senthil}},
  \bibinfo{journal}{Physical Review Research} \textbf{\bibinfo{volume}{1}}
  (\bibinfo{year}{2019}), ISSN \bibinfo{issn}{2643-1564},
  \urlprefix\url{http://dx.doi.org/10.1103/PhysRevResearch.1.033126}.

\bibitem[{\citenamefont{Chen et~al.}(2020)\citenamefont{Chen, Sharpe, Fox,
  Zhang, Wang, Jiang, Lyu, Li, Watanabe, Taniguchi et~al.}}]{hall}
\bibinfo{author}{\bibfnamefont{G.}~\bibnamefont{Chen}},
  \bibinfo{author}{\bibfnamefont{A.~L.} \bibnamefont{Sharpe}},
  \bibinfo{author}{\bibfnamefont{E.~J.} \bibnamefont{Fox}},
  \bibinfo{author}{\bibfnamefont{Y.-H.} \bibnamefont{Zhang}},
  \bibinfo{author}{\bibfnamefont{S.}~\bibnamefont{Wang}},
  \bibinfo{author}{\bibfnamefont{L.}~\bibnamefont{Jiang}},
  \bibinfo{author}{\bibfnamefont{B.}~\bibnamefont{Lyu}},
  \bibinfo{author}{\bibfnamefont{H.}~\bibnamefont{Li}},
  \bibinfo{author}{\bibfnamefont{K.}~\bibnamefont{Watanabe}},
  \bibinfo{author}{\bibfnamefont{T.}~\bibnamefont{Taniguchi}},
  \bibnamefont{et~al.}, \bibinfo{journal}{Nature}
  \textbf{\bibinfo{volume}{579}}, \bibinfo{pages}{56每61}
  (\bibinfo{year}{2020}), ISSN \bibinfo{issn}{1476-4687},
  \urlprefix\url{http://dx.doi.org/10.1038/s41586-020-2049-7}.

\bibitem[{\citenamefont{Serlin et~al.}(2019)\citenamefont{Serlin, Tschirhart,
  Polshyn, Zhang, Zhu, Watanabe, Taniguchi, Balents, and Young}}]{hall2}
\bibinfo{author}{\bibfnamefont{M.}~\bibnamefont{Serlin}},
  \bibinfo{author}{\bibfnamefont{C.~L.} \bibnamefont{Tschirhart}},
  \bibinfo{author}{\bibfnamefont{H.}~\bibnamefont{Polshyn}},
  \bibinfo{author}{\bibfnamefont{Y.}~\bibnamefont{Zhang}},
  \bibinfo{author}{\bibfnamefont{J.}~\bibnamefont{Zhu}},
  \bibinfo{author}{\bibfnamefont{K.}~\bibnamefont{Watanabe}},
  \bibinfo{author}{\bibfnamefont{T.}~\bibnamefont{Taniguchi}},
  \bibinfo{author}{\bibfnamefont{L.}~\bibnamefont{Balents}}, \bibnamefont{and}
  \bibinfo{author}{\bibfnamefont{A.~F.} \bibnamefont{Young}},
  \bibinfo{journal}{Science} \textbf{\bibinfo{volume}{367}},
  \bibinfo{pages}{900每903} (\bibinfo{year}{2019}), ISSN
  \bibinfo{issn}{1095-9203},
  \urlprefix\url{http://dx.doi.org/10.1126/science.aay5533}.

\bibitem[{\citenamefont{Cao et~al.}(2020)\citenamefont{Cao, Chowdhury,
  Rodan-Legrain, Rubies-Bigorda, Watanabe, Taniguchi, Senthil, and
  Jarillo-Herrero}}]{moirestrange}
\bibinfo{author}{\bibfnamefont{Y.}~\bibnamefont{Cao}},
  \bibinfo{author}{\bibfnamefont{D.}~\bibnamefont{Chowdhury}},
  \bibinfo{author}{\bibfnamefont{D.}~\bibnamefont{Rodan-Legrain}},
  \bibinfo{author}{\bibfnamefont{O.}~\bibnamefont{Rubies-Bigorda}},
  \bibinfo{author}{\bibfnamefont{K.}~\bibnamefont{Watanabe}},
  \bibinfo{author}{\bibfnamefont{T.}~\bibnamefont{Taniguchi}},
  \bibinfo{author}{\bibfnamefont{T.}~\bibnamefont{Senthil}}, \bibnamefont{and}
  \bibinfo{author}{\bibfnamefont{P.}~\bibnamefont{Jarillo-Herrero}},
  \bibinfo{journal}{Physical Review Letters} \textbf{\bibinfo{volume}{124}}
  (\bibinfo{year}{2020}), ISSN \bibinfo{issn}{1079-7114},
  \urlprefix\url{http://dx.doi.org/10.1103/PhysRevLett.124.076801}.

\bibitem[{\citenamefont{Polshyn et~al.}(2019)\citenamefont{Polshyn, Yankowitz,
  Chen, Zhang, Watanabe, Taniguchi, Dean, and Young}}]{youngstrange}
\bibinfo{author}{\bibfnamefont{H.}~\bibnamefont{Polshyn}},
  \bibinfo{author}{\bibfnamefont{M.}~\bibnamefont{Yankowitz}},
  \bibinfo{author}{\bibfnamefont{S.}~\bibnamefont{Chen}},
  \bibinfo{author}{\bibfnamefont{Y.}~\bibnamefont{Zhang}},
  \bibinfo{author}{\bibfnamefont{K.}~\bibnamefont{Watanabe}},
  \bibinfo{author}{\bibfnamefont{T.}~\bibnamefont{Taniguchi}},
  \bibinfo{author}{\bibfnamefont{C.~R.} \bibnamefont{Dean}}, \bibnamefont{and}
  \bibinfo{author}{\bibfnamefont{A.~F.} \bibnamefont{Young}},
  \bibinfo{journal}{Nature Physics} \textbf{\bibinfo{volume}{15}},
  \bibinfo{pages}{1011每1016} (\bibinfo{year}{2019}), ISSN
  \bibinfo{issn}{1745-2481},
  \urlprefix\url{http://dx.doi.org/10.1038/s41567-019-0596-3}.

\bibitem[{\citenamefont{Jarillo-Herrero}(2019)}]{pablotalk}
\bibinfo{author}{\bibfnamefont{P.}~\bibnamefont{Jarillo-Herrero}},
  \emph{\bibinfo{title}{{Magic Angle Graphene Transport Phenomenology}}},
  \bibinfo{howpublished}{\url{http://online.kitp.ucsb.edu/online/bands_m19/jar%
illoherrero/}} (\bibinfo{year}{2019}).

\bibitem[{\citenamefont{{Cao} et~al.}(2020)\citenamefont{{Cao},
  {Rodan-Legrain}, {Park}, {Noah Yuan}, {Watanabe}, {Taniguchi}, {Fernandes},
  {Fu}, and {Jarillo-Herrero}}}]{pablonematic}
\bibinfo{author}{\bibfnamefont{Y.}~\bibnamefont{{Cao}}},
  \bibinfo{author}{\bibfnamefont{D.}~\bibnamefont{{Rodan-Legrain}}},
  \bibinfo{author}{\bibfnamefont{J.~M.} \bibnamefont{{Park}}},
  \bibinfo{author}{\bibfnamefont{F.}~\bibnamefont{{Noah Yuan}}},
  \bibinfo{author}{\bibfnamefont{K.}~\bibnamefont{{Watanabe}}},
  \bibinfo{author}{\bibfnamefont{T.}~\bibnamefont{{Taniguchi}}},
  \bibinfo{author}{\bibfnamefont{R.~M.} \bibnamefont{{Fernandes}}},
  \bibinfo{author}{\bibfnamefont{L.}~\bibnamefont{{Fu}}}, \bibnamefont{and}
  \bibinfo{author}{\bibfnamefont{P.}~\bibnamefont{{Jarillo-Herrero}}},
  \bibinfo{journal}{arXiv e-prints} \bibinfo{eid}{arXiv:2004.04148}
  (\bibinfo{year}{2020}), \eprint{2004.04148}.

\bibitem[{\citenamefont{Lee et~al.}(2019)\citenamefont{Lee, Khalaf, Liu, Liu,
  Hao, Kim, and Vishwanath}}]{TDBGt}
\bibinfo{author}{\bibfnamefont{J.~Y.} \bibnamefont{Lee}},
  \bibinfo{author}{\bibfnamefont{E.}~\bibnamefont{Khalaf}},
  \bibinfo{author}{\bibfnamefont{S.}~\bibnamefont{Liu}},
  \bibinfo{author}{\bibfnamefont{X.}~\bibnamefont{Liu}},
  \bibinfo{author}{\bibfnamefont{Z.}~\bibnamefont{Hao}},
  \bibinfo{author}{\bibfnamefont{P.}~\bibnamefont{Kim}}, \bibnamefont{and}
  \bibinfo{author}{\bibfnamefont{A.}~\bibnamefont{Vishwanath}},
  \bibinfo{journal}{Nature Communications} \textbf{\bibinfo{volume}{10}}
  (\bibinfo{year}{2019}), ISSN \bibinfo{issn}{2041-1723},
  \urlprefix\url{http://dx.doi.org/10.1038/s41467-019-12981-1}.

\bibitem[{\citenamefont{Dodaro et~al.}(2018)\citenamefont{Dodaro, Kivelson,
  Schattner, Sun, and Wang}}]{kivelsontbg}
\bibinfo{author}{\bibfnamefont{J.~F.} \bibnamefont{Dodaro}},
  \bibinfo{author}{\bibfnamefont{S.~A.} \bibnamefont{Kivelson}},
  \bibinfo{author}{\bibfnamefont{Y.}~\bibnamefont{Schattner}},
  \bibinfo{author}{\bibfnamefont{X.~Q.} \bibnamefont{Sun}}, \bibnamefont{and}
  \bibinfo{author}{\bibfnamefont{C.}~\bibnamefont{Wang}},
  \bibinfo{journal}{Phys. Rev. B} \textbf{\bibinfo{volume}{98}},
  \bibinfo{pages}{075154} (\bibinfo{year}{2018}),
  \urlprefix\url{https://link.aps.org/doi/10.1103/PhysRevB.98.075154}.

\bibitem[{\citenamefont{Po et~al.}(2018)\citenamefont{Po, Zou, Vishwanath, and
  Senthil}}]{senthil}
\bibinfo{author}{\bibfnamefont{H.~C.} \bibnamefont{Po}},
  \bibinfo{author}{\bibfnamefont{L.}~\bibnamefont{Zou}},
  \bibinfo{author}{\bibfnamefont{A.}~\bibnamefont{Vishwanath}},
  \bibnamefont{and} \bibinfo{author}{\bibfnamefont{T.}~\bibnamefont{Senthil}},
  \bibinfo{journal}{Phys. Rev. X} \textbf{\bibinfo{volume}{8}},
  \bibinfo{pages}{031089} (\bibinfo{year}{2018}),
  \urlprefix\url{https://link.aps.org/doi/10.1103/PhysRevX.8.031089}.

\bibitem[{\citenamefont{Zou et~al.}(2018)\citenamefont{Zou, Po, Vishwanath, and
  Senthil}}]{band1}
\bibinfo{author}{\bibfnamefont{L.}~\bibnamefont{Zou}},
  \bibinfo{author}{\bibfnamefont{H.~C.} \bibnamefont{Po}},
  \bibinfo{author}{\bibfnamefont{A.}~\bibnamefont{Vishwanath}},
  \bibnamefont{and} \bibinfo{author}{\bibfnamefont{T.}~\bibnamefont{Senthil}},
  \bibinfo{journal}{Phys. Rev. B} \textbf{\bibinfo{volume}{98}},
  \bibinfo{pages}{085435} (\bibinfo{year}{2018}),
  \urlprefix\url{https://link.aps.org/doi/10.1103/PhysRevB.98.085435}.

\bibitem[{\citenamefont{Zhang and Senthil}(2019)}]{band4}
\bibinfo{author}{\bibfnamefont{Y.-H.} \bibnamefont{Zhang}} \bibnamefont{and}
  \bibinfo{author}{\bibfnamefont{T.}~\bibnamefont{Senthil}},
  \bibinfo{journal}{Physical Review B} \textbf{\bibinfo{volume}{99}}
  (\bibinfo{year}{2019}), ISSN \bibinfo{issn}{2469-9969},
  \urlprefix\url{http://dx.doi.org/10.1103/PhysRevB.99.205150}.

\bibitem[{\citenamefont{Oganesyan et~al.}(2001)\citenamefont{Oganesyan,
  Kivelson, and Fradkin}}]{fradkinnematic}
\bibinfo{author}{\bibfnamefont{V.}~\bibnamefont{Oganesyan}},
  \bibinfo{author}{\bibfnamefont{S.~A.} \bibnamefont{Kivelson}},
  \bibnamefont{and} \bibinfo{author}{\bibfnamefont{E.}~\bibnamefont{Fradkin}},
  \bibinfo{journal}{Phys. Rev. B} \textbf{\bibinfo{volume}{64}},
  \bibinfo{pages}{195109} (\bibinfo{year}{2001}),
  \urlprefix\url{https://link.aps.org/doi/10.1103/PhysRevB.64.195109}.

\bibitem[{\citenamefont{Fradkin et~al.}(2010)\citenamefont{Fradkin, Kivelson,
  Lawler, Eisenstein, and Mackenzie}}]{nematicreview}
\bibinfo{author}{\bibfnamefont{E.}~\bibnamefont{Fradkin}},
  \bibinfo{author}{\bibfnamefont{S.~A.} \bibnamefont{Kivelson}},
  \bibinfo{author}{\bibfnamefont{M.~J.} \bibnamefont{Lawler}},
  \bibinfo{author}{\bibfnamefont{J.~P.} \bibnamefont{Eisenstein}},
  \bibnamefont{and} \bibinfo{author}{\bibfnamefont{A.~P.}
  \bibnamefont{Mackenzie}}, \bibinfo{journal}{Annual Review of Condensed Matter
  Physics} \textbf{\bibinfo{volume}{1}}, \bibinfo{pages}{153}
  (\bibinfo{year}{2010}),
  \eprint{https://doi.org/10.1146/annurev-conmatphys-070909-103925},
  \urlprefix\url{https://doi.org/10.1146/annurev-conmatphys-070909-103925}.

\bibitem[{\citenamefont{Hertz}(1976)}]{hertz}
\bibinfo{author}{\bibfnamefont{J.~A.} \bibnamefont{Hertz}},
  \bibinfo{journal}{Phys. Rev. B} \textbf{\bibinfo{volume}{14}},
  \bibinfo{pages}{1165} (\bibinfo{year}{1976}),
  \urlprefix\url{https://link.aps.org/doi/10.1103/PhysRevB.14.1165}.

\bibitem[{\citenamefont{Millis}(1993)}]{millis}
\bibinfo{author}{\bibfnamefont{A.~J.} \bibnamefont{Millis}},
  \bibinfo{journal}{Phys. Rev. B} \textbf{\bibinfo{volume}{48}},
  \bibinfo{pages}{7183} (\bibinfo{year}{1993}),
  \urlprefix\url{https://link.aps.org/doi/10.1103/PhysRevB.48.7183}.

\bibitem[{\citenamefont{L\"ohneysen et~al.}(2007)\citenamefont{L\"ohneysen,
  Rosch, Vojta, and W\"olfle}}]{review}
\bibinfo{author}{\bibfnamefont{H.~v.} \bibnamefont{L\"ohneysen}},
  \bibinfo{author}{\bibfnamefont{A.}~\bibnamefont{Rosch}},
  \bibinfo{author}{\bibfnamefont{M.}~\bibnamefont{Vojta}}, \bibnamefont{and}
  \bibinfo{author}{\bibfnamefont{P.}~\bibnamefont{W\"olfle}},
  \bibinfo{journal}{Rev. Mod. Phys.} \textbf{\bibinfo{volume}{79}},
  \bibinfo{pages}{1015} (\bibinfo{year}{2007}),
  \urlprefix\url{https://link.aps.org/doi/10.1103/RevModPhys.79.1015}.

\bibitem[{\citenamefont{Wu}(1982)}]{clock}
\bibinfo{author}{\bibfnamefont{F.~Y.} \bibnamefont{Wu}}, \bibinfo{journal}{Rev.
  Mod. Phys.} \textbf{\bibinfo{volume}{54}}, \bibinfo{pages}{235}
  (\bibinfo{year}{1982}),
  \urlprefix\url{https://link.aps.org/doi/10.1103/RevModPhys.54.235}.

\bibitem[{\citenamefont{Fucito and Parisi}(1981)}]{6epsilon}
\bibinfo{author}{\bibfnamefont{F.}~\bibnamefont{Fucito}} \bibnamefont{and}
  \bibinfo{author}{\bibfnamefont{G.}~\bibnamefont{Parisi}},
  \bibinfo{journal}{J. Phys. A: Math. Gen.} \textbf{\bibinfo{volume}{14}},
  \bibinfo{pages}{L507} (\bibinfo{year}{1981}).

\bibitem[{\citenamefont{Bi et~al.}(2020)\citenamefont{Bi, Lake, and
  Senthil}}]{bisenthil2}
\bibinfo{author}{\bibfnamefont{Z.}~\bibnamefont{Bi}},
  \bibinfo{author}{\bibfnamefont{E.}~\bibnamefont{Lake}}, \bibnamefont{and}
  \bibinfo{author}{\bibfnamefont{T.}~\bibnamefont{Senthil}},
  \bibinfo{journal}{Physical Review Research} \textbf{\bibinfo{volume}{2}}
  (\bibinfo{year}{2020}), ISSN \bibinfo{issn}{2643-1564},
  \urlprefix\url{http://dx.doi.org/10.1103/PhysRevResearch.2.023031}.

\bibitem[{\citenamefont{Li et~al.}(2017)\citenamefont{Li, Jiang, Jian, and
  Yao}}]{yao2017}
\bibinfo{author}{\bibfnamefont{Z.-X.} \bibnamefont{Li}},
  \bibinfo{author}{\bibfnamefont{Y.-F.} \bibnamefont{Jiang}},
  \bibinfo{author}{\bibfnamefont{S.-K.} \bibnamefont{Jian}}, \bibnamefont{and}
  \bibinfo{author}{\bibfnamefont{H.}~\bibnamefont{Yao}},
  \bibinfo{journal}{Nature Communications} \textbf{\bibinfo{volume}{8}}
  (\bibinfo{year}{2017}), ISSN \bibinfo{issn}{2041-1723},
  \urlprefix\url{http://dx.doi.org/10.1038/s41467-017-00167-6}.

\bibitem[{\citenamefont{Nayak and Wilczek}(1994{\natexlab{a}})}]{nayak1}
\bibinfo{author}{\bibfnamefont{C.}~\bibnamefont{Nayak}} \bibnamefont{and}
  \bibinfo{author}{\bibfnamefont{F.}~\bibnamefont{Wilczek}},
  \bibinfo{journal}{Nuclear Physics B} \textbf{\bibinfo{volume}{417}},
  \bibinfo{pages}{359 } (\bibinfo{year}{1994}{\natexlab{a}}), ISSN
  \bibinfo{issn}{0550-3213},
  \urlprefix\url{http://www.sciencedirect.com/science/article/pii/055032139490%
4774}.

\bibitem[{\citenamefont{Nayak and Wilczek}(1994{\natexlab{b}})}]{nayak2}
\bibinfo{author}{\bibfnamefont{C.}~\bibnamefont{Nayak}} \bibnamefont{and}
  \bibinfo{author}{\bibfnamefont{F.}~\bibnamefont{Wilczek}},
  \bibinfo{journal}{Nuclear Physics B} \textbf{\bibinfo{volume}{430}},
  \bibinfo{pages}{534 } (\bibinfo{year}{1994}{\natexlab{b}}), ISSN
  \bibinfo{issn}{0550-3213},
  \urlprefix\url{http://www.sciencedirect.com/science/article/pii/055032139490%
1589}.

\bibitem[{\citenamefont{Mross et~al.}(2010)\citenamefont{Mross, McGreevy, Liu,
  and Senthil}}]{mrossnfl}
\bibinfo{author}{\bibfnamefont{D.~F.} \bibnamefont{Mross}},
  \bibinfo{author}{\bibfnamefont{J.}~\bibnamefont{McGreevy}},
  \bibinfo{author}{\bibfnamefont{H.}~\bibnamefont{Liu}}, \bibnamefont{and}
  \bibinfo{author}{\bibfnamefont{T.}~\bibnamefont{Senthil}},
  \bibinfo{journal}{Phys. Rev. B} \textbf{\bibinfo{volume}{82}},
  \bibinfo{pages}{045121} (\bibinfo{year}{2010}),
  \urlprefix\url{https://link.aps.org/doi/10.1103/PhysRevB.82.045121}.

\bibitem[{\citenamefont{Lee}(2009)}]{sslee}
\bibinfo{author}{\bibfnamefont{S.-S.} \bibnamefont{Lee}},
  \bibinfo{journal}{Phys. Rev. B} \textbf{\bibinfo{volume}{80}},
  \bibinfo{pages}{165102} (\bibinfo{year}{2009}),
  \urlprefix\url{https://link.aps.org/doi/10.1103/PhysRevB.80.165102}.

\bibitem[{\citenamefont{Metlitski and Sachdev}(2010)}]{maxnematic}
\bibinfo{author}{\bibfnamefont{M.~A.} \bibnamefont{Metlitski}}
  \bibnamefont{and} \bibinfo{author}{\bibfnamefont{S.}~\bibnamefont{Sachdev}},
  \bibinfo{journal}{Phys. Rev. B} \textbf{\bibinfo{volume}{82}},
  \bibinfo{pages}{075127} (\bibinfo{year}{2010}),
  \urlprefix\url{https://link.aps.org/doi/10.1103/PhysRevB.82.075127}.

\bibitem[{\citenamefont{Varma et~al.}(1989)\citenamefont{Varma, Littlewood,
  Schmitt-Rink, Abrahams, and Ruckenstein}}]{mfl}
\bibinfo{author}{\bibfnamefont{C.~M.} \bibnamefont{Varma}},
  \bibinfo{author}{\bibfnamefont{P.~B.} \bibnamefont{Littlewood}},
  \bibinfo{author}{\bibfnamefont{S.}~\bibnamefont{Schmitt-Rink}},
  \bibinfo{author}{\bibfnamefont{E.}~\bibnamefont{Abrahams}}, \bibnamefont{and}
  \bibinfo{author}{\bibfnamefont{A.~E.} \bibnamefont{Ruckenstein}},
  \bibinfo{journal}{Phys. Rev. Lett.} \textbf{\bibinfo{volume}{63}},
  \bibinfo{pages}{1996} (\bibinfo{year}{1989}),
  \urlprefix\url{https://link.aps.org/doi/10.1103/PhysRevLett.63.1996}.

\bibitem[{\citenamefont{{Xu} et~al.}(2020)\citenamefont{{Xu}, {Klein}, {Sun},
  {Chubukov}, and {Meng}}}]{meng23}
\bibinfo{author}{\bibfnamefont{X.~Y.} \bibnamefont{{Xu}}},
  \bibinfo{author}{\bibfnamefont{A.}~\bibnamefont{{Klein}}},
  \bibinfo{author}{\bibfnamefont{K.}~\bibnamefont{{Sun}}},
  \bibinfo{author}{\bibfnamefont{A.~V.} \bibnamefont{{Chubukov}}},
  \bibnamefont{and} \bibinfo{author}{\bibfnamefont{Z.~Y.}
  \bibnamefont{{Meng}}}, \bibinfo{journal}{arXiv e-prints}
  \bibinfo{eid}{arXiv:2003.11573} (\bibinfo{year}{2020}), \eprint{2003.11573}.

\bibitem[{\citenamefont{Xie et~al.}(2019)\citenamefont{Xie, Lian, J\"{a}ck,
  Liu, Chiu, Watanabe, Taniguchi, Bernevig, and Yazdani}}]{strain}
\bibinfo{author}{\bibfnamefont{Y.}~\bibnamefont{Xie}},
  \bibinfo{author}{\bibfnamefont{B.}~\bibnamefont{Lian}},
  \bibinfo{author}{\bibfnamefont{B.}~\bibnamefont{J\"{a}ck}},
  \bibinfo{author}{\bibfnamefont{X.}~\bibnamefont{Liu}},
  \bibinfo{author}{\bibfnamefont{C.-L.} \bibnamefont{Chiu}},
  \bibinfo{author}{\bibfnamefont{K.}~\bibnamefont{Watanabe}},
  \bibinfo{author}{\bibfnamefont{T.}~\bibnamefont{Taniguchi}},
  \bibinfo{author}{\bibfnamefont{B.~A.} \bibnamefont{Bernevig}},
  \bibnamefont{and} \bibinfo{author}{\bibfnamefont{A.}~\bibnamefont{Yazdani}},
  \bibinfo{journal}{Nature} \textbf{\bibinfo{volume}{572}},
  \bibinfo{pages}{101} (\bibinfo{year}{2019}).

\bibitem[{\citenamefont{Bi et~al.}(2019)\citenamefont{Bi, Yuan, and
  Fu}}]{strainband}
\bibinfo{author}{\bibfnamefont{Z.}~\bibnamefont{Bi}},
  \bibinfo{author}{\bibfnamefont{N.~F.~Q.} \bibnamefont{Yuan}},
  \bibnamefont{and} \bibinfo{author}{\bibfnamefont{L.}~\bibnamefont{Fu}},
  \bibinfo{journal}{Phys. Rev. B} \textbf{\bibinfo{volume}{100}},
  \bibinfo{pages}{035448} (\bibinfo{year}{2019}),
  \urlprefix\url{https://link.aps.org/doi/10.1103/PhysRevB.100.035448}.

\bibitem[{\citenamefont{Schattner et~al.}(2016)\citenamefont{Schattner,
  Lederer, Kivelson, and Berg}}]{nematicnumerics1}
\bibinfo{author}{\bibfnamefont{Y.}~\bibnamefont{Schattner}},
  \bibinfo{author}{\bibfnamefont{S.}~\bibnamefont{Lederer}},
  \bibinfo{author}{\bibfnamefont{S.~A.} \bibnamefont{Kivelson}},
  \bibnamefont{and} \bibinfo{author}{\bibfnamefont{E.}~\bibnamefont{Berg}},
  \bibinfo{journal}{Phys. Rev. X} \textbf{\bibinfo{volume}{6}},
  \bibinfo{pages}{031028} (\bibinfo{year}{2016}),
  \urlprefix\url{https://link.aps.org/doi/10.1103/PhysRevX.6.031028}.

\bibitem[{\citenamefont{Lederer et~al.}(2017)\citenamefont{Lederer, Schattner,
  Berg, and Kivelson}}]{nematicnumerics2}
\bibinfo{author}{\bibfnamefont{S.}~\bibnamefont{Lederer}},
  \bibinfo{author}{\bibfnamefont{Y.}~\bibnamefont{Schattner}},
  \bibinfo{author}{\bibfnamefont{E.}~\bibnamefont{Berg}}, \bibnamefont{and}
  \bibinfo{author}{\bibfnamefont{S.~A.} \bibnamefont{Kivelson}},
  \bibinfo{journal}{Proceedings of the National Academy of Sciences}
  \textbf{\bibinfo{volume}{114}}, \bibinfo{pages}{4905} (\bibinfo{year}{2017}),
  ISSN \bibinfo{issn}{0027-8424},
  \eprint{https://www.pnas.org/content/114/19/4905.full.pdf},
  \urlprefix\url{https://www.pnas.org/content/114/19/4905}.

\bibitem[{\citenamefont{Xu et~al.}(2019)\citenamefont{Xu, Hong~Liu, Pan, Qi,
  Sun, and Meng}}]{nematicnumerics3}
\bibinfo{author}{\bibfnamefont{X.~Y.} \bibnamefont{Xu}},
  \bibinfo{author}{\bibfnamefont{Z.}~\bibnamefont{Hong~Liu}},
  \bibinfo{author}{\bibfnamefont{G.}~\bibnamefont{Pan}},
  \bibinfo{author}{\bibfnamefont{Y.}~\bibnamefont{Qi}},
  \bibinfo{author}{\bibfnamefont{K.}~\bibnamefont{Sun}}, \bibnamefont{and}
  \bibinfo{author}{\bibfnamefont{Z.~Y.} \bibnamefont{Meng}},
  \bibinfo{journal}{Journal of Physics: Condensed Matter}
  \textbf{\bibinfo{volume}{31}}, \bibinfo{pages}{463001}
  (\bibinfo{year}{2019}), ISSN \bibinfo{issn}{1361-648X},
  \urlprefix\url{http://dx.doi.org/10.1088/1361-648X/ab3295}.

\bibitem[{\citenamefont{Metlitski et~al.}(2015)\citenamefont{Metlitski, Mross,
  Sachdev, and Senthil}}]{nflpair1}
\bibinfo{author}{\bibfnamefont{M.~A.} \bibnamefont{Metlitski}},
  \bibinfo{author}{\bibfnamefont{D.~F.} \bibnamefont{Mross}},
  \bibinfo{author}{\bibfnamefont{S.}~\bibnamefont{Sachdev}}, \bibnamefont{and}
  \bibinfo{author}{\bibfnamefont{T.}~\bibnamefont{Senthil}},
  \bibinfo{journal}{Phys. Rev. B} \textbf{\bibinfo{volume}{91}},
  \bibinfo{pages}{115111} (\bibinfo{year}{2015}),
  \urlprefix\url{https://link.aps.org/doi/10.1103/PhysRevB.91.115111}.

\bibitem[{\citenamefont{Lederer et~al.}(2015)\citenamefont{Lederer, Schattner,
  Berg, and Kivelson}}]{nflpair2}
\bibinfo{author}{\bibfnamefont{S.}~\bibnamefont{Lederer}},
  \bibinfo{author}{\bibfnamefont{Y.}~\bibnamefont{Schattner}},
  \bibinfo{author}{\bibfnamefont{E.}~\bibnamefont{Berg}}, \bibnamefont{and}
  \bibinfo{author}{\bibfnamefont{S.~A.} \bibnamefont{Kivelson}},
  \bibinfo{journal}{Phys. Rev. Lett.} \textbf{\bibinfo{volume}{114}},
  \bibinfo{pages}{097001} (\bibinfo{year}{2015}),
  \urlprefix\url{https://link.aps.org/doi/10.1103/PhysRevLett.114.097001}.

\bibitem[{\citenamefont{Rech et~al.}(2006)\citenamefont{Rech, P\'epin, and
  Chubukov}}]{nflpair3}
\bibinfo{author}{\bibfnamefont{J.}~\bibnamefont{Rech}},
  \bibinfo{author}{\bibfnamefont{C.}~\bibnamefont{P\'epin}}, \bibnamefont{and}
  \bibinfo{author}{\bibfnamefont{A.~V.} \bibnamefont{Chubukov}},
  \bibinfo{journal}{Phys. Rev. B} \textbf{\bibinfo{volume}{74}},
  \bibinfo{pages}{195126} (\bibinfo{year}{2006}),
  \urlprefix\url{https://link.aps.org/doi/10.1103/PhysRevB.74.195126}.

\bibitem[{\citenamefont{Chubukov et~al.}(2004)\citenamefont{Chubukov, P\'epin,
  and Rech}}]{nflinstable}
\bibinfo{author}{\bibfnamefont{A.~V.} \bibnamefont{Chubukov}},
  \bibinfo{author}{\bibfnamefont{C.}~\bibnamefont{P\'epin}}, \bibnamefont{and}
  \bibinfo{author}{\bibfnamefont{J.}~\bibnamefont{Rech}},
  \bibinfo{journal}{Phys. Rev. Lett.} \textbf{\bibinfo{volume}{92}},
  \bibinfo{pages}{147003} (\bibinfo{year}{2004}),
  \urlprefix\url{https://link.aps.org/doi/10.1103/PhysRevLett.92.147003}.

\bibitem[{\citenamefont{Wu et~al.}(2019)\citenamefont{Wu, Hwang, and
  Das~Sarma}}]{phononstrange}
\bibinfo{author}{\bibfnamefont{F.}~\bibnamefont{Wu}},
  \bibinfo{author}{\bibfnamefont{E.}~\bibnamefont{Hwang}}, \bibnamefont{and}
  \bibinfo{author}{\bibfnamefont{S.}~\bibnamefont{Das~Sarma}},
  \bibinfo{journal}{Phys. Rev. B} \textbf{\bibinfo{volume}{99}},
  \bibinfo{pages}{165112} (\bibinfo{year}{2019}),
  \urlprefix\url{https://link.aps.org/doi/10.1103/PhysRevB.99.165112}.

\end{thebibliography}

\end{document}